\documentclass{INTERSPEECH2023}
\usepackage{cite}
\usepackage{amsfonts}
\usepackage{wrapfig}
\usepackage{setspace}
\usepackage{multirow}
\usepackage{multicol}
\usepackage{algorithm}
\usepackage{enumitem}
\usepackage{subcaption}
\usepackage{mathtools}
\usepackage{mathalpha}
\usepackage{amsmath}
\usepackage{booktabs}       
\usepackage{cite}
\interspeechcameraready

\title{Adversarial Learning of Intermediate Acoustic Feature \\ for End-to-End Lightweight Text-to-Speech}
\name{Hyungchan Yoon$^1$, Seyun Um$^1$, Changhwan Kim$^1$, Hong-Goo Kang$^1$}
\address{
  $^1$Department of Electrical and Electronic Engineering, Yonsei University, Seoul, Korea}
\email{[hcy71, chkim, syum]@dsp.yonsei.ac.kr, hgkang@yonsei.ac.kr}

\begin{document}

\maketitle
 
\begin{abstract}
To simplify the generation process, several text-to-speech (TTS) systems implicitly learn intermediate latent representations instead of relying on predefined features (e.g., mel-spectrogram). However, their generation quality is unsatisfactory as these representations lack speech variances.
In this paper, we improve TTS performance by adding \emph{prosody embeddings} to the latent representations. During training, we extract reference prosody embeddings from mel-spectrograms, and during inference, we estimate these embeddings from text using generative adversarial networks (GANs). Using GANs, we reliably estimate the prosody embeddings in a fast way, which have complex distributions due to the dynamic nature of speech. We also show that the prosody embeddings work as efficient features for learning a robust alignment between text and acoustic features. Our proposed model surpasses several publicly available models with less parameters and computational complexity in comparative experiments.
\end{abstract}

\vspace{-2pt}
\section{Introduction}

The advancement of neural networks has led to significant quality improvement in text-to-speech (TTS) systems.
Depending on the structural differences when converting text into intermediate features (e.g., mel-spectrogram), neural TTS models can be classified as either autoregressive (AR) or non-autoregressive (non-AR).
AR-based methods \cite{wang2017tacotron,shen2018natural,Li2019NeuralSS} successfully generate high-quality speech but have low inference speed because they cannot be implemented in a parallel manner.
To solve this problem, various types of non-AR TTS models have been proposed \cite{Ren2019FastSpeechFR,ren2020fastspeech,kim2020glow,elias2021parallel}.

Recently, non-AR TTS models have been extended to a single-stage end-to-end structure that incorporates a generative model-based vocoder into the training process \cite{ren2020fastspeech} (FastSpeech 2s), \cite{donahue2020end,nguyen2021litetts}.
These models train entire text-to-waveform conversion chains under a unified framework. Thus, they can utilize a latent representation that is trained to implicitly contain acoustic information
instead of being constrained to predefined intermediate features such as the mel-spectrogram.
However, the synthesized speech quality of such models is unsatisfactory because providing sufficient speech variance information needed for direct text-to-waveform mapping is difficult without the process of generating a mel-spectrogram.

In this paper, we propose AILTTS, a single-stage lightweight TTS model that achieves high performance by effectively providing speech variance information required for direct text-to-waveform mapping.
A key feature that represents speech variance is the prosody-related acoustic feature. To extract such features, we adopt a prosody encoder (posterior) that uses a mel-spectrogram as an input. We denote the output of the prosody encoder as a reference prosody embedding, which is conditioned for the text-to-waveform conversion process during training.
Because the input of the prosody encoder, i.e., mel-spectrogram, is not available during inference,
we adopt a prosody predictor (prior) that estimates the aforementioned reference prosody embedding only from a text input.
To enhance the estimation power by using a generative model, we apply generative adversarial networks (GANs) to the prosody predictor.
The proposed prosody predictor requires few parameters and simplifies the inference process while exhibiting high output performance compared to conventional methods. 
In addition, the proposed method is helpful in enhancing the performance of the likelihood-based time alignment that inherently finds the timing information between text and mel-spectrogram.
\begin{figure*}[t!]
    \centering
    \includegraphics[width=0.99\textwidth]{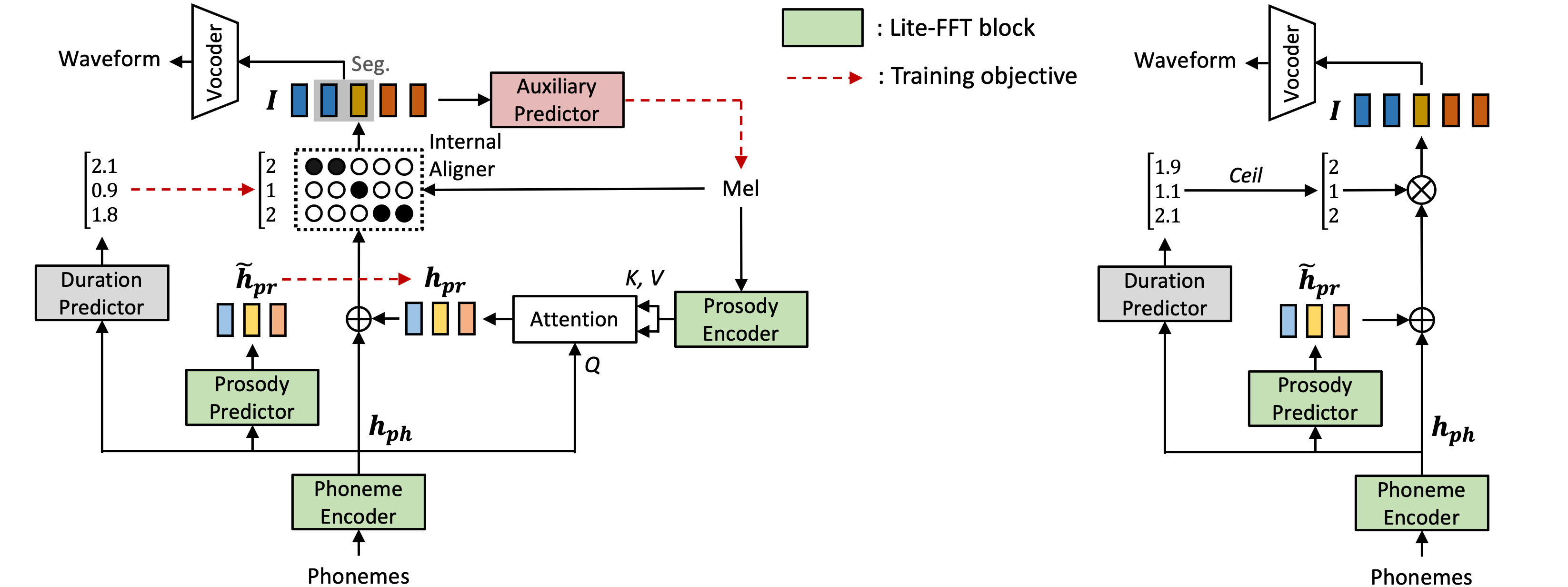}
    \caption{\normalfont{Overall architecture of the training (left) and inference (right) stages of the proposed model. The green box represents the Lite-FFT block, a lightweight transformer-based encoder. The red arrow represents the training objective. The conditional discriminator for the prosody predictor is omitted for brevity.}}
    \label{fig:model}
\vspace{-5pt}
\end{figure*}
Our main contributions are as follows.
\begin{itemize}
    \setlength\itemsep{0.5em}
    \item We effectively provide the speech variance in single-stage TTS system by conditioning the prosody-related acoustic embedding. In particular, we adopt an adversarial training to make the prior reliably estimate the reference prosody embedding from text input. 
    \item  As we fully utilize the characteristics of the prosody embedding, the internal alignment model converges fast in the early stage of training as well as attains robust performance in the end.
    \item Experiments confirm that the proposed method considerably enhances the quality of generated speech even with a small number of parameters\footnote{The generated audio samples for the experiments are also available at our demo page \normalfont{\url{https://hcy71o.github.io/AILTTS\_demo/}}}.
\end{itemize}


\vspace{-2pt}
\section{Related Work}
\label{Realted Work}
\vspace{-2pt}
\subsection{Single-stage TTS}
\vspace{-1pt}
There have been several attempts to implement a single-stage TTS system that can utilize a trainable latent representation. FastSpeech 2s \cite{ren2020fastspeech} and EATS \cite{donahue2020end} successfully implemented the idea by directly connecting the text encoder and the
publicly accessible vocoder \cite{yamamoto2020parallel, binkowski2019high}, but their generation quality was not on par with two-stage models. VITS \cite{kim2021conditional} adopts a variational inference with normalizing flows to connect a feature extractor \cite{kim2020glow} and a vocoder \cite{kong2020hifi}, and significantly outperforms two-stage models. However, the normalizing flow module \cite{oord2016wavenet} leads to an additional computational complexity. LiteTTS \cite{nguyen2021litetts}, which is our backbone model in this work, requires fewer parameters and low computational complexity with small memory footprints; however, its output speech quality is unsatisfactory because its prior’s estimation power is insufficient.
To solve this problem by strengthening the role of the prior, we adopt a conditional discriminator and an adversarial training.

\vspace{-2pt}
\subsection{Alignment between text and acoustic features}
To align the sequence of text and acoustic features, i.e., mel-spectrogram, FastSpeech 2/2s \cite{ren2020fastspeech} utilized an external aligner \cite{mcauliffe2017montreal} that extracts the time-duration of each input text. Recently, likelihood-based aligners \cite{kim2020glow,badlani2022one} have been proposed, which remove the reliance on external aligners by internally estimating the time duration. In this work, we introduce a likelihood-based internal aligner conditioned on the prosody embedding mentioned previously, which robustly and quickly performs alignments compared to the conventional methods.
\vspace{-2pt}
\section{Method}
\label{Method}
\vspace{-2pt}
\subsection{Overview}
\vspace{-1pt}
Figure \ref{fig:model} presents a block diagram of the proposed model built using the LiteTTS \cite{nguyen2021litetts} baseline. 
Our model consists of a phoneme encoder, a prosody encoder (posterior), a prosody predictor (prior), an internal aligner including a duration predictor, an auxiliary predictor, and a vocoder.
The overall training process is as follows.
We first calculate the phoneme-scale prosody embedding $h_{pr}$ taking an attention with the prosody encoder output (key and value) and a phoneme encoder output $h_{ph}$ (query).
Then, we time-align the joint embeddings $h_{ph}+h_{pr}$ to mel-spectrogram using the internal aligner, where the aligned embedding is denoted as $I$. Finally, we generate a waveform using a vocoder by conditioning the aligned embedding. 
To build a low complexity architecture, we adopt a lightweight transformer-based encoder \cite{nguyen2021litetts} for the phoneme encoder, the prosody encoder, and the prosody predictor.
\vspace{-2pt}

\subsection{Prosody predictor with conditional discriminator}
\begin{figure}[t!]
    \centerline{\includegraphics[width=.47\textwidth]{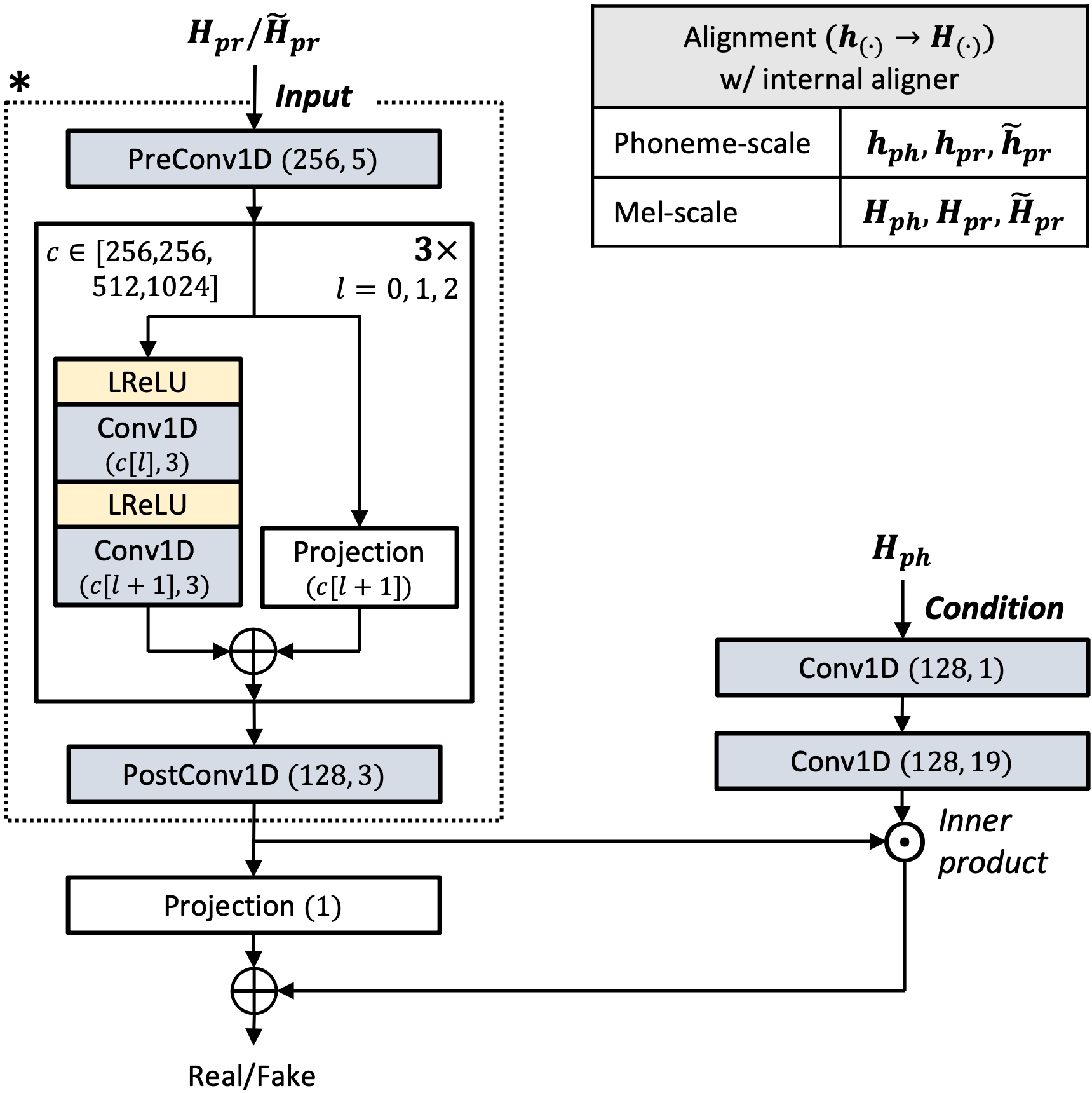}}
    \caption{\normalfont{Architecture of the proposed conditional discriminator and the auxiliary predictor*. The main and conditioning branches comprising 1D convolutional layers are designed to have the same receptive field size.}}
    \label{fig:discriminator}
    \vspace{-10pt}
\end{figure}

The main purpose of the proposed prosody predictor is to predict the target prosody embedding $h_{pr}$ from the input phonetic embedding $h_{ph}$. 
Considering the dynamic nature of the prosody embedding, we leverage an effective generative model architecture that includes various types of discriminators. Defining the prosody predictor as a generator, the proposed discriminator distinguishes the target prosody embedding $h_{pr}$ from the predicted embedding $\tilde{h}_{pr}$, and utilizes phonetic information as a condition. We adopt a projection-based conditional discriminator \cite{miyato2018cgans} that uses the phonetic embedding $h_{ph}$ as a condition (shown in Fig. \ref{fig:discriminator}).
In addition, we apply feature matching loss \cite{larsen2016autoencoding} between the generated and target feature maps to stabilize the GAN-based training process, where feature maps are defined as the outputs of all the 1D convolution layers prior to the $PostConv1D$ layer. In particular, we extract seven feature maps: one from the $PreConv1D$ layer and six from three residual 1D convolutional blocks.

Meanwhile, considering the characteristics of our model, we use two additional tricks while designing the discriminator. 
In the voice-generation process, the $I$ that is aligned to the timing information of the \textit{mel-spectrogram}, is used as the input to the vocoder. Accordingly, we first design the discriminator to distinguish between two prosody embeddings in the
aligned domain instead of the original phoneme domain.
In the former part of the discriminator, we align the the time scale of phoneme-wise embeddings to that of the mel-spectrogram by utilizing duration values estimated from the internal aligner.
Second, we design the discriminator to have the same receptive field size as the vocoder, which is extremely small due to a GPU memory constraint in the training process.
It enables the discriminator to efficiently capture the diverse patterns of input prosody embeddings.
Based on the generative and discriminative loss of a least-squares GAN \cite{Mao2017LeastSG}, we define the total prosody predictor loss by considering both reconstruction loss $\mathcal{L}_{recon}$ and feature matching loss $\mathcal{L}_{fm}$:
\begin{equation}
\mathcal{L}_{G}= \mathbb{E}_{(\tilde{{H}}_{pr},{H}_{ph})}[(D(\tilde{H}_{pr},H_{ph}
)-1)^2 ] + \mathcal{L}_{recon} + \mathcal{L}_{fm},
\end{equation}
\begin{equation}
\mathcal{L}_{D}=
\mathbb{E}_{({H}_{pr},\tilde{{H}}_{pr},{H}_{ph})} [(D({H}_{pr},{H}_{ph})-1)^2\nonumber\\+(D(\tilde{{H}}_{pr},{H}_{ph}))^2],
\end{equation}
\begin{equation}
\mathcal{L}_{recon}= ||\tilde{{H}}_{pr}- {H}_{pr}||_{1},\, \mathcal{L}_{fm}=\sum_{i=1}^{7}||\tilde{F}^{i}_{pr}-F^{i}_{pr}||_{1},
\end{equation}
where $H_{(\cdot)}$ denotes the embedding mapped into a time scale of the mel-spectrogram, and $F^{i}$ denotes the $i$-th feature map of the discriminator.

\subsection{Prosody-conditioned internal aligner}
To learn the time alignment between the phoneme and mel-spectrogram without using an external aligner, we adopt a likelihood-based internal aligner \cite{kim2020glow,badlani2022one}. As specified in \cite{badlani2022one}, we maximize the likelihood of monotonic alignments from a probability matrix calculated by the $L2$-distance between two encoded features.
Then, phoneme durations (binary matrix) are obtained by selecting the most probable path from the probability matrix. The gap between the two matrices is reduced by minimizing their KL-divergence.

The proposed aligner utilizes the joint embeddings $h_{ph}$ + $h_{pr}$ for the phonetic feature and the mel-spectrogram for the acoustic feature. Since $h_{pr}$ itself contains local acoustic information mapped into the phoneme-level by a previously applied attention module, learning the alignment becomes much easier for the aligner than only using  $h_{ph}$. Following \cite{ren2020fastspeech}, we jointly train the duration predictor, which accepts the phonetic embedding $h_{ph}$ with a stop gradient.
As a result, the proposed aligner increases the accuracy of alignment, enabling the duration predictor to estimate more accurate durations.

\subsection{Final training loss}
The total loss for training our model is defined as follows:
\begin{equation}
\mathcal{L}_{total}= \mathcal{L}_{var}+\mathcal{L}_{align}+\mathcal{L}_{pred}+\mathcal{L}_{voc}+\mathcal{L}_{aux},
\end{equation}
where $\mathcal{L}_{var}$ includes pitch and energy prediction losses applied to the output of the prosody encoder in \cite{nguyen2021litetts}, and $\mathcal{L}_{align}$ includes loss functions for the internal aligner\cite{badlani2022one}, including the duration predictor. $\mathcal{L}_{pred}$ and $\mathcal{L}_{voc}$ represent total losses from the prosody predictor and the vocoder, respectively, in accordance with their GAN structure. $\mathcal{L}_{aux}$ is defined by the $L1$ loss between the target mel-spectrogram and the predicted mel-spectrogram obtained by the output of the auxiliary predictor. 

\textbf{Auxiliary predictor}
To further provide acoustic information to the aligned embedding $I$, we adopt the auxiliary predictor whose input is $I$.
Its architecture is identical to that marked `*' in Fig. \ref{fig:discriminator}, with the following exceptions: 1) The number of output channels for the $PostConv1D$ layer is set to the dimension of the mel-spectrogram; 2) Layer normalization is applied to the last stage of every residual 1D convolutional block. 
Note that its receptive field is also the same as that of the vocoder; It enables the auxiliary predictor to efficiently provide acoustic information to the vocoder input $I$. In addition, as the auxiliary predictor is only used for the training stage, the number of parameters and computational complexity are not increased.
\vspace{-2pt}
\section{Experiments}
\label{Experiments}
\vspace{-2pt}

\subsection{Experimental setup}
\noindent\textbf{Dataset and Model Settings} We evaluated our model on the LJSpeech dataset \cite{ljspeech17}. We converted text sequences into phoneme sequences with an open-source tool\footnote{\normalfont{\url{https://github.com/Kyubyong/g2p}}}. To calculate spectrograms, we set the FFT size, window size, hop size to 1024, 1024, and 256, respectively, regarding the sampling rate of 22050Hz. Mel-spectrograms were obtained by applying an 80-band mel filter bank. We adopted Univnet-c16 \cite{jang2021univnet} as a vocoder, which has beneficial lightweight properties by using a location-variable convolution (LVC) technique \cite{Zeng2021LVCNetEC}. The dimensions of all hidden embeddings were set to 256, and the receptive field of the vocoder, auxiliary predictor, and the conditional discriminator was set to 19.

\begin{table*}[t!]
	\centering
	\caption{\normalfont{Comparisons of MOS with 95\% confidence intervals, number of parameters, inference speed, and CER. 15 test samples were used for MOS test, and 80 test samples were used for inference speed and CER tests.}}\label{table:MOS}
        \renewcommand{\arraystretch}{0.95}
    	\begin{tabular}{cccccc}
            \toprule
            \multicolumn{1}{l}{\textbf{Model}} & \multicolumn{1}{c}{\textbf{MOS}} & \multicolumn{1}{c}{\textbf{\#Params.}} & \multicolumn{1}{c}{\textbf{CPU inference}}&\multicolumn{1}{c}{\textbf{Real-time}}&\multicolumn{1}{c}{\textbf{CER}}
            \\ \midrule
            \multicolumn{1}{l}{Ground truth} & $4.80\pm0.07$ &-&-&-&3.99\\
            \multicolumn{1}{l}{AILTTS} & $3.93\pm0.11$ & $13.4$M & $329.3$kHz& $\times14.94$&9.18\\
            \multicolumn{1}{l}{LiteTTS} & $3.71\pm0.10$ &$13.4$M& $329.8$kHz& $\times14.95$&5.22\\
            \multicolumn{1}{l}{Tacotron 2} & $3.93\pm0.10$ &$29.4$M& $41.8$kHz&$\times1.90$&11.04\\
            \multicolumn{1}{l}{Glow-TTS} & $3.57\pm0.12$ &$30.0$M& $208.6$kHz&$\times9.46$&6.89\\
            \bottomrule
    	\end{tabular}
\end{table*}

\noindent\textbf{Training Configuration and Inference}
We trained our model with the Adam \cite{Kingma2015AdamAM} optimizer with $\beta_{1}=0.8, \beta_{2}=0.99$. We set the learning rate and its decaying factor to $2e-4$ and $0.999$, respectively. The batch size was set to $24$.
The entire model was trained in a fully end-to-end manner, except for the prosody and duration predictors.
Because their training objectives (${h}_{pr}$ and durations) do not converge to meaningful values during the early training stage, these two modules are jointly trained with other modules after 300k steps. For the prosody predictor, we applied only reconstruction loss for the first 50k steps (from 300k to 350k steps) to stabilize the training of GANs. During inference, the prosody embedding and duration values predicted by the two abovementioned modules were used as shown in the right side of Fig \ref{fig:model}.

\noindent\textbf{Models for Comparison}
We compared our model with both single-stage and two-stage models.  To clearly demonstrate the effectiveness of our method, we
maintained the vocoder unified for all implemented models used for experiments.
We first experimented with LiteTTS by replacing the HiFi-GAN V3 \cite{kong2020hifi} vocoder with the UnivNet-c16 vocoder for a fair comparison.
For the two-stage models, we adopted two popular AR and non-AR TTS models (feature extractors) as baselines and UnivNet-c16 as the vocoder. Tacotron 2\cite{shen2018natural} and Glow-TTS\cite{kim2020glow} were used as AR- and a non-AR-type feature extractors, respectively. Here, we utilized pre-trained weights for the feature extractors\footnote{Tacotron 2:\url{https://github.com/NVIDIA/tacotron2}\\Glow-TTS:\url{https://github.com/jaywalnut310/glow-tts}}, and trained the vocoder\footnote{UnivNet-c16:\url{https://github.com/mindslab-ai/univnet}} using a 80-dim mel-spectrogram as input.

\subsection{Results}
Table \ref{table:MOS} summarizes the evaluation metrics for our proposed AILTTS model, including mean opinion scores (MOS), model parameters, inference speed, and character error rate (CER).
For MOS, we randomly selected 15 test audio samples to measure the naturalness of synthesized speech.
To measure the intelligibility, we calculated CER of the synthesized speech by transcribing it using a pre-trained speech recognition model from the SpeechBrain toolkit \cite{ravanelli2021speechbrain}. 
For measuring the CER and inference speed, we generated 80 samples using arbitrary text scripts as input.

In terms of naturalness, our AILTTS model outperformed two non-AR baselines, LiteTTS (by +0.22 MOS) and Glow-TTS (by +0.36 MOS), with a small number of parameters ($13.4$M). Also, our model produced natural speech comparable to that of the AR-based TTS model, Tacotron 2. 
By contrast, our model achieves slightly higher (worse) CER than the most similar baseline, LiteTTS, showing the trade-off between naturalness and intelligibility.
It can be explained in terms of the mode-seeking behavior of GANs. In the first viewpoint, AILTTS effectively addresses the problem of prosody oversmoothing in LiteTTS by modeling the prosody embedding with a more complex distribution with GANs\footnote{Note that LiteTTS only uses L1 loss for the prior (i.e., prosody predictor), which models the prosody embedding with a simple Laplacian distribution.}.
However, it can introduce some difficulties in robustly predicting the prosody embedding, which lead to lower intelligibility in our experiments. In summary, AILTTS produces speech that sounds natural than LiteTTS, albeit with slightly lower intelligibility score.
By discussing the impact of GAN's mode-seeking behavior on the distribution of the prosody embedding, our findings provide a more complete understanding of the naturalness-intelligibility trade-off observed in AILTTS, emphasizing its effectiveness in improving synthesized speech quality. 
Furthermore, we evaluated the inference speed using an Intel Core i5 Quad-Core 2.0-GHz CPU and found that our proposed model performs almost as fast as LiteTTS, while outperforming the baseline two-stage models.

\begin{table}[t]
	\begin{center}
    	\caption{\normalfont{CMOS results of ablation studies.}}\label{table:CMOS}
        \begingroup
	    \renewcommand{\arraystretch}{0.95}
    	\begin{tabular}{cc}
            \toprule
            \multicolumn{1}{l}{\textbf{Model}} & \multicolumn{1}{l}{\textbf{CMOS}}
            \\ \midrule
            \multicolumn{1}{l}{AILTTS} & $0$ \\ \midrule
            \multicolumn{1}{l}{without Conditional Discriminator} & $-0.214$ \\
            \multicolumn{1}{l}{without Prosody-conditioned Aligner} & $-0.871$ \\ \bottomrule
    	\end{tabular}
    	\endgroup
	\end{center}
 \vspace{-20pt}
\end{table}

\begin{figure}
    \centerline{\includegraphics[width=0.5\textwidth]{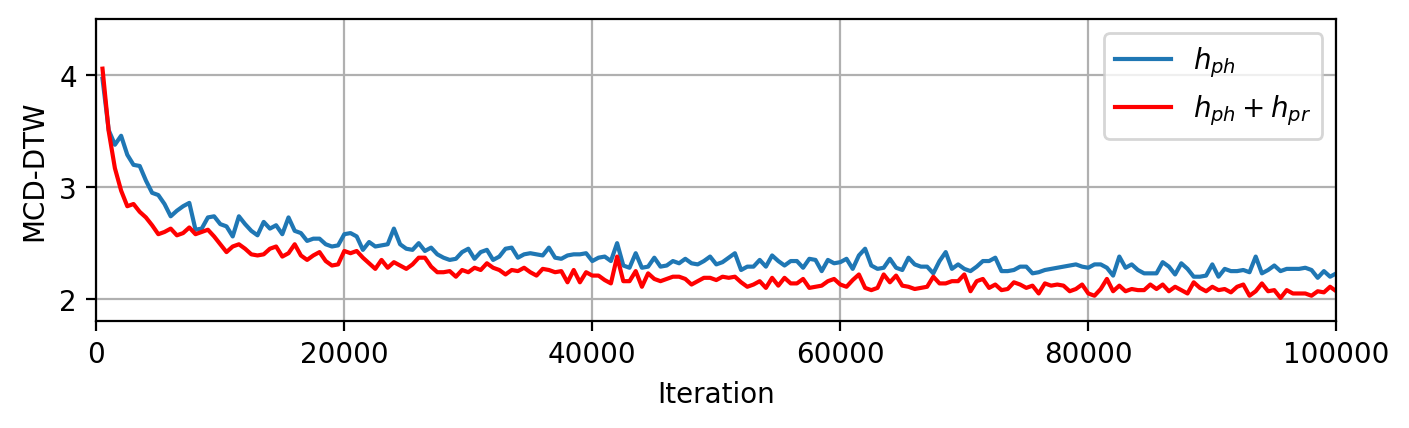}}
    \caption{\normalfont{Comparison of the convergence rate between a conventional method (using ${h}_{ph}$ as the aligner input) and the proposed method (using ${h}_{ph}+{h}_{pr}$ as the aligner input). For each iteration, the MCD-DTW values from 10 test samples were averaged.}}
\vspace{-13pt}
    \label{fig:MCD}
\end{figure}

\subsection{Analysis}
\noindent\textbf{Ablation Study}
We performed a comparison MOS (CMOS) test on the test dataset to investigate the effect of the proposed methods on the overall performance. 
Table \ref{table:CMOS} presents the results of the two ablation studies.
In the first study, the conditional discriminator was removed and only reconstruction loss was applied.
The result shows that adopting a GANs for the prosody predictor enhances the naturalness of generated speech because it improves the estimation power of the prior compared to the one using only L1 loss.
For the second study, the input of the internal aligner ${h}_{ph}+{h}_{pr}$ was replaced with phonetic embedding ${h}_{ph}$. The experimental result also demonstrates that providing phoneme-level acoustic information (prosody embedding) to the aligner is helpful for further improving the generation quality due to the robust time alignment between text and acoustic features.

\noindent\textbf{Alignment Convergence} We also compared the alignment convergence rate of the two experiments from the ablation study of the prosody-conditioned aligner. To check this, we calculated the mean mel-cepstral distance (MCD) between synthesized and ground-truth audio samples after synchronizing them using time with dynamic time warping (DTW) \cite{battenberg2020location}. 
As shown in Fig \ref{fig:MCD}, the MCD-DTW values of the proposed aligner rapidly decrease during the early stage of training (prior to 4k iterations). Further, the proposed method outperformed the baseline with lower MCD-DTW values in every iteration, in accordance with the second result of the ablation study.
\vspace{-2pt}
\section{Conclusion}
\vspace{-2pt}
\label{Conclusion}
In this paper, we propose an adversarial learning method to reliably estimate a prosody embedding, an intermediate acoustic feature that plays a significant role in generating natural speech in single-stage lightweight TTS systems.
The proposed model, equipped with an internal aligner that synergizes with the prosody embedding, outperforms the Glow-TTS based model (+0.36 MOS) and LiteTTS (+0.22 MOS) in terms of naturalness while preserving its lightweight properties, at the cost of a minor decrease in intelligibility.
\vspace{-2pt}
\section{Acknowledgement}
\vspace{-2pt}
This work was supported by Voice\&Avatar, NAVER Cloud, Seongnam, Korea.
\clearpage

\bibliographystyle{IEEEtran}
\bibliography{mybib}

\begin{thebibliography}{10}
\providecommand{\url}[1]{#1}
\csname url@samestyle\endcsname
\providecommand{\newblock}{\relax}
\providecommand{\bibinfo}[2]{#2}
\providecommand{\BIBentrySTDinterwordspacing}{\spaceskip=0pt\relax}
\providecommand{\BIBentryALTinterwordstretchfactor}{4}
\providecommand{\BIBentryALTinterwordspacing}{\spaceskip=\fontdimen2\font plus
\BIBentryALTinterwordstretchfactor\fontdimen3\font minus
  \fontdimen4\font\relax}
\providecommand{\BIBforeignlanguage}[2]{{%
\expandafter\ifx\csname l@#1\endcsname\relax
\typeout{** WARNING: IEEEtran.bst: No hyphenation pattern has been}%
\typeout{** loaded for the language `#1'. Using the pattern for}%
\typeout{** the default language instead.}%
\else
\language=\csname l@#1\endcsname
\fi
#2}}
\providecommand{\BIBdecl}{\relax}
\BIBdecl

\bibitem{wang2017tacotron}
Y.~Wang, R.~Skerry-Ryan, D.~Stanton, Y.~Wu, R.~J. Weiss, N.~Jaitly, Z.~Yang,
  Y.~Xiao, Z.~Chen, S.~Bengio \emph{et~al.}, ``Tacotron: Towards end-to-end
  speech synthesis,'' in \emph{Proc. INTERSPEECH}, 2017, pp. 4006--4010.

\bibitem{shen2018natural}
J.~Shen, R.~Pang, R.~J. Weiss, M.~Schuster, N.~Jaitly, Z.~Yang, Z.~Chen,
  Y.~Zhang, Y.~Wang, R.~Skerrv-Ryan \emph{et~al.}, ``Natural {TTS} synthesis by
  conditioning wavenet on mel spectrogram predictions,'' in \emph{Proc.
  ICASSP}, 2018, pp. 4779--4783.

\bibitem{Li2019NeuralSS}
N.~Li, S.~Liu, Y.~Liu, S.~Zhao, and M.~Liu, ``Neural speech synthesis with
  transformer network,'' in \emph{Proc. AAAI}, 2019, p. 6706–6713.

\bibitem{Ren2019FastSpeechFR}
Y.~Ren, Y.~Ruan, X.~Tan, T.~Qin, S.~Zhao, Z.~Zhao, and T.-Y. Liu, ``Fastspeech:
  Fast, robust and controllable text-to-speech,'' in \emph{Proc. NIPS}, 2019,
  p. 3171–3180.

\bibitem{ren2020fastspeech}
Y.~Ren, C.~Hu, X.~Tan, T.~Qin, S.~Zhao, Z.~Zhao, and T.-Y. Liu, ``Fastspeech 2:
  Fast and high-quality end-to-end text-to-speech,'' in \emph{Proc. ICLR},
  2021.

\bibitem{kim2020glow}
J.~Kim, S.~Kim, J.~Kong, and S.~Yoon, ``Glow-{TTS}: A generative flow for
  text-to-speech via monotonic alignment search,'' in \emph{Proc. NIPS}, 2020,
  p. 8067–8077.

\bibitem{elias2021parallel}
I.~Elias, H.~Zen, J.~Shen, Y.~Zhang, Y.~Jia, R.~J. Weiss, and Y.~Wu, ``Parallel
  tacotron: Non-autoregressive and controllable {TTS},'' in \emph{Proc.
  ICASSP}, 2021, pp. 5709--5713.

\bibitem{donahue2020end}
J.~Donahue, S.~Dieleman, M.~Binkowski, E.~Elsen, and K.~Simonyan, ``End-to-end
  adversarial text-to-speech,'' in \emph{Proc. ICLR}, 2021.

\bibitem{nguyen2021litetts}
H.-K. Nguyen, K.~Jeong, S.~Um, M.-J. Hwang, E.~Song, and H.-G. Kang,
  ``Lite{TTS}: A lightweight mel-spectrogram-free text-to-wave synthesizer
  based on generative adversarial networks,'' in \emph{Proc. INTERSPEECH},
  2021, pp. 3595--3599.

\bibitem{yamamoto2020parallel}
R.~Yamamoto, E.~Song, and J.-M. Kim, ``Parallel wavegan: A fast waveform
  generation model based on generative adversarial networks with
  multi-resolution spectrogram,'' in \emph{Proc. ICASSP}, 2020, pp. 6199--6203.

\bibitem{binkowski2019high}
M.~Bi{\'n}kowski, J.~Donahue, S.~Dieleman, A.~Clark, E.~Elsen, N.~Casagrande,
  L.~C. Cobo, and K.~Simonyan, ``High fidelity speech synthesis with
  adversarial networks,'' in \emph{Proc. ICLR}, 2020.

\bibitem{kim2021conditional}
J.~Kim, J.~Kong, and J.~Son, ``Conditional variational autoencoder with
  adversarial learning for end-to-end text-to-speech,'' in \emph{Proc. ICML},
  2021, pp. 5530--5540.

\bibitem{kong2020hifi}
J.~Kong, J.~Kim, and J.~Bae, ``{Hifi-GAN}: Generative adversarial networks for
  efficient and high fidelity speech synthesis,'' in \emph{Proc. NIPS}, 2020,
  pp. 17\,022--17\,033.

\bibitem{oord2016wavenet}
A.~v.~d. Oord, S.~Dieleman, H.~Zen, K.~Simonyan, O.~Vinyals, A.~Graves,
  N.~Kalchbrenner, A.~Senior, and K.~Kavukcuoglu, ``Wavenet: A generative model
  for raw audio,'' \emph{arXiv preprint arXiv:1609.03499}, 2016.

\bibitem{mcauliffe2017montreal}
M.~McAuliffe, M.~Socolof, S.~Mihuc, M.~Wagner, and M.~Sonderegger, ``Montreal
  forced aligner: Trainable text-speech alignment using kaldi.'' in \emph{Proc.
  INTERSPEECH}, 2017, pp. 498--502.

\bibitem{badlani2022one}
R.~Badlani, A.~{\L}a{\'n}cucki, K.~J. Shih, R.~Valle, W.~Ping, and
  B.~Catanzaro, ``One {TTS} alignment to rule them all,'' in \emph{Proc.
  ICASSP}, 2022, pp. 6092--6096.

\bibitem{miyato2018cgans}
T.~Miyato and M.~Koyama, ``c{GAN}s with projection discriminator,'' in
  \emph{Proc. ICLR}, 2018.

\bibitem{larsen2016autoencoding}
A.~B.~L. Larsen, S.~K. S{\o}nderby, H.~Larochelle, and O.~Winther,
  ``Autoencoding beyond pixels using a learned similarity metric,'' in
  \emph{Proc. ICML}, 2016, p. 1558–1566.

\bibitem{Mao2017LeastSG}
X.~Mao, Q.~Li, H.~Xie, R.~Y.~K. Lau, Z.~Wang, and S.~P. Smolley, ``Least
  squares generative adversarial networks,'' in \emph{Proc. ICCV}, 2017, pp.
  2813--2821.

\bibitem{ljspeech17}
K.~Ito and L.~Johnson, ``The lj speech dataset,''
  \url{https://keithito.com/LJ-Speech-Dataset/}, 2017.

\bibitem{jang2021univnet}
W.~Jang, D.~Lim, J.~Yoon, B.~Kim, and J.~Kim, ``Univnet: A neural vocoder with
  multi-resolution spectrogram discriminators for high-fidelity waveform
  generation,'' in \emph{Proc. INTERSPEECH}, 2021, pp. 2207--2211.

\bibitem{Zeng2021LVCNetEC}
Z.~Zeng, J.~Wang, N.~Cheng, and J.~Xiao, ``Lvcnet: Efficient
  condition-dependent modeling network for waveform generation,'' in
  \emph{Proc. ICASSP}, 2021, pp. 6054--6058.

\bibitem{Kingma2015AdamAM}
D.~P. Kingma and J.~Ba, ``Adam: A method for stochastic optimization,'' in
  \emph{Proc. ICLR}, 2015.

\bibitem{ravanelli2021speechbrain}
M.~Ravanelli, T.~Parcollet, P.~Plantinga, A.~Rouhe, S.~Cornell, L.~Lugosch,
  C.~Subakan, N.~Dawalatabad, A.~Heba, J.~Zhong \emph{et~al.}, ``Speechbrain: A
  general-purpose speech toolkit,'' \emph{arXiv preprint arXiv:2106.04624},
  2021.

\bibitem{battenberg2020location}
E.~Battenberg, R.~Skerry-Ryan, S.~Mariooryad, D.~Stanton, D.~Kao, M.~Shannon,
  and T.~Bagby, ``Location-relative attention mechanisms for robust long-form
  speech synthesis,'' in \emph{Proc. ICASSP}, 2020, pp. 6194--6198.

\end{thebibliography}

\end{document}